\documentstyle[aps,prb,twocolumn,eqsecnum,epsfig,floats]{revtex}

\begin{document}

\draft
\preprint{}

\twocolumn[\hsize\textwidth\columnwidth\hsize\csname @twocolumnfalse\endcsname

\title{Andreev-Tunneling, Coulomb Blockade, and \\Resonant Transport of
Non-Local
Spin-Entangled Electrons}
\author{Patrik Recher, Eugene V. Sukhorukov, and Daniel Loss}
\address{
Department of Physics and Astronomy, University of Basel,\\
Klingelbergstrasse 82, CH-4056 Basel, Switzerland}

\date{\today}

\maketitle
\begin{abstract}
We propose and  analyze a spin-entangler for electrons based on an s-wave
superconductor coupled to two  quantum dots 
each of which is tunnel-coupled to normal Fermi leads. 
 We show that in the presence of a voltage bias and in the Coulomb
blockade regime  two correlated electrons provided by the Andreev process
 can coherently tunnel from the superconductor via different dots into different
leads. The spin-singlet coming from the Cooper pair
remains preserved in this process, and the setup provides a source of 
mobile and nonlocal spin-entangled electrons.
The transport current is calculated and shown to be dominated by a
two-particle
Breit-Wigner resonance which allows the injection of  two spin-entangled
electrons
into different leads at
exactly the same orbital energy, which is a crucial requirement for
the detection of spin entanglement via noise measurements.
The coherent tunneling of
both electrons into the same lead is suppressed by the on-site Coulomb
repulsion and/or the superconducting gap, while the tunneling into different
leads is suppressed through the initial separation of the tunneling electrons.
In the regime of interest the particle-hole excitations of the leads are shown
to be negligible.
The Aharonov-Bohm oscillations in the current are  shown
to contain single- and two-electron periods with amplitudes that both vanish 
with increasing Coulomb repulsion albeit differently fast.

\end{abstract}
\vskip2pc]
\narrowtext

\section{Introduction}
The creation of nonlocal pairwise-entangled quantum states,
so-called Einstein-Podolsky-Rosen (EPR) pairs \cite{Einstein}, is essential
for secure quantum communication \cite{Bennett84}, dense coding
and quantum teleportation \cite{BennettNature}, or more
fundamental, for testing  the violation of Bell's inequality \cite{Bell}.
Such tests already exist for photons but not yet for {\it massive}
particles such as electrons since it is difficult to produce and to detect
entangled electrons. However, 
there is strong experimental evidence that electron spins in a semiconductor
environment show unusually long dephasing times approaching microseconds and
that they can be transported phase coherently over distances exceeding 100$\mu m$
\cite{{Kikkawa1,Kikkawa2,Awschalom,Fiederling,Ohno}}. This
makes spins of electrons in semiconductors promising candidates
for carriers of quantum information (qubits)\cite{Loss97,BEL}. 
In particular, we have recently proposed a
setup\cite{BLS} consisting of a spin-entangler and a beam splitter 
where the spin-entanglement is detectable 
via electronic transport properties.
We have shown that
 the current-current correlations (noise) is enhanced
if the entangled electrons are spin-singlets leading to bunching behavior,
whereas the noise is suppressed  for spin-triplets leading to
antibunching behavior.

In Ref.\cite{BLS} we assumed the existence of  an entangler, i.e. a device
 that generates spin-singlets which are made out of two
electrons which reside in different but degenerate orbital states,
and we focussed on the question of how to detect spin-entangled electrons
via transport and noise measurements. Here, we address  the problem of how
to implement
such an entangler in a solid state device.
We have found\cite{BLS} that for such noise measurements, which are based on
two-particle interference effects, it is
absolutely crucial that both  electrons, coming from
different
leads, possess the
{\em same orbital energy}. If the orbital energies of the two entangled
electrons are
different, the electrons cannot interfere with each other,
and thus 
spin correlations would not be observable in the noise\cite{BLS}.

\begin{figure}[h]
\centerline{\psfig{file=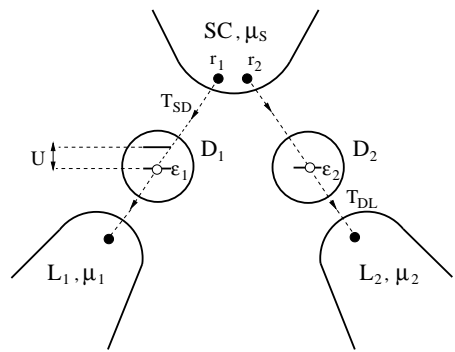,width=6cm}}
\vspace{5mm}
\label{spintabl}
\caption{The entangler setup: Two spin-entangled electrons forming a Cooper pair
can tunnel with amplitude $T_{SD}$ from points ${\bf r}_{1}$ and ${\bf
r}_{2}$ of the superconductor, SC, to two dots, $D_{1}$ and $D_{2}$, by means
of Andreev tunneling. The dots are tunnel-coupled to normal  leads ${\rm L_{1}}$
and ${\rm L_{2}}$, with tunneling amplitude $T_{DL}$. 
The superconductor and leads are kept at chemical potentials $\mu_{S}$ and
$\mu_{l}$, resp.}
\end{figure}

In the following we propose a setup  which involves a superconductor coupled to
two quantum dots which themselves are coupled to normal leads, see Fig.\ 1. 
 We show that such a setup acts as an entangler which meets all the requirements
needed for a successful detection
of spin entanglement via noise measurements.  
In  previous work\cite{CBL}
 we showed that in equilibrium the spin correlations of an s-wave
superconductor
induce a spin-singlet state between two electrons each of which
resides on a separate
quantum dot which both are weakly coupled to the same superconductor
(but not among themselves). This
non-local
spin-entanglement leads then to observable effects in a generalized
Josephson
junction setup\cite{CBL}. In the present work we 
consider a
{\em non-equilibrium} situation where an applied voltage bias drives a
stationary
current of pairwise spin-entangled electrons from the superconductor
through the
quantum dots into the leads, see
Figs.\ 1 and 2.

\section{Qualitative Description of the Andreev Entangler}

We begin with a qualitative description of the entangler and 
its principal mechanism based on Andreev processes and Coulomb blockade
effects. In subsequent sections we introduce then the  Hamiltonian and calculate the
stationary current in detail. 
We consider an s-wave superconductor which acts as a natural
source of spin-entangled electrons, since
the electrons form Cooper pairs with singlet
spin-wavefunctions\cite{Schrieffer}.
The superconductor, which is  held at the chemical potential $\mu_{S}$,
is weakly
coupled by tunnel barriers to two separate quantum dots $D_{1}$ and
$D_{2}$ which themselves are weakly
coupled to Fermi liquid leads $L_{1}$ and $L_{2}$, resp., both held at the
same chemical potential $\mu_{1}=\mu_{2}$. 
The corresponding tunneling amplitudes
between superconductor and dots, and dots-leads, are denoted
by $T_{SD}$ and $T_{DL}$, resp. (for simplicity we assume them to be equal
for both dots and leads).

In general, the tunnel-coupling of a superconductor to
a normal region allows for coherent transport of two electrons of opposite
spins due
to Andreev tunneling\cite{Schrieffer}, while single-electron
tunneling is suppressed\cite{Glazman}.
In the present setup, we envision a situation where
the two electrons are forced to
tunnel coherently into {\em different} leads rather than both into the same
lead.
This situation can be enforced in the presence of two intermediate
quantum dots which
are assumed to be in
the Coulomb blockade regime\cite{Kouwenhoven} so that the state with the two
electrons
being on the same quantum dot is strongly suppressed, and thus the electrons
will preferably tunnel into separate dots and subsequently into separate
leads (this will be
quantified in the following).

By applying a bias
voltage $\Delta\mu=\mu_{S}-\mu_{l}>0$ transport of entangled electrons
occurs
from the superconductor via the dots to the leads.
The chemical potentials $\epsilon_{1}$ and $\epsilon_{2}$
of the quantum dots can be tuned by external gate
voltages\cite{Kouwenhoven} such that
the coherent tunneling of two electrons into
different leads is at resonance, described by a two-particle Breit-Wigner
resonance peaked at
$\epsilon_{1}=\epsilon_{2}=\mu_{S}$. In contrast, we will see that the
current for the coherent
tunneling of two electrons into the {\it same} lead is suppressed by the on-site
Coulomb $U$
repulsion of a
quantum dot and/or by the superconducting gap $\Delta$.

Next, we introduce  the relevant parameters describing the proposed
device and specify
their regime of interest.
First we note that to
avoid unwanted correlations with electrons already on the quantum dots
 one could work
in the cotunneling regime\cite{Kouwenhoven} where the number of
electrons on the dots
are fixed and the resonant levels $\epsilon_{l}$, $l=1,2$ cannot be
occupied. However, we
prefer to work at the resonances $\epsilon_{l}\simeq\mu_{S}$, since then the total
current
and the desired suppression of tunneling into the same lead is maximized in
this regime. Also, the desired injection of the two electrons into different
leads but at the {\it same orbital energy} is then achieved.
It turns out to be most convenient to work in the regime where the dot levels
$\epsilon_{l}$ have vanishing occupation probability. For this purpose
we require that the dot-lead coupling is much stronger than the
superconductor-dot coupling, i.e. $|T_{SD}|< |T_{DL}|$, so that
electrons
which enter the dots from the superconductor will leave the quantum
dots to
the leads much faster than new electrons can be provided from the
superconductor. In addition, a
stationary occupation due to the coupling to the leads is 
exponentially small
if $\Delta\mu>k_{B}T$, $T$ being the temperature and $k_{B}$ the Boltzmann
constant.
Thus in this asymmetric barrier case, the resonant dot levels $\epsilon_{l}$ can
be occupied only during a virtual process.

Next, the quantum dots in the ground state are allowed to contain an arbitrary but even
number of electrons, $N_{D}=\rm{even}$, with total spin zero
(i.e. antiferromagnetic filling of the dots).
An odd number $N_D$ must be excluded since a simple
spin-flip on the quantum dot would be possible in the transport process
and as a result the desired entanglement would be lost.
Further, we have to make
sure that also spin flip processes of the following kind are excluded.
Consider
an electron that tunnels from the superconductor into a given dot. Now, it
is
possible in principle (e.g. in a sequential tunneling
process\cite{Kouwenhoven})
that another electron
with the opposite spin leaves the dot and tunnels into the lead, and, again,
the
desired entanglement would be lost. However, such spin flip processes will
be
excluded if the energy level spacing of the quantum dots, $\delta\epsilon$,
(assumed
to be similar for both dots) exceeds both, temperature $k_{B}T$ and bias
voltage
$\Delta\mu$.
A serious source of
entanglement-loss is given by electron hole-pair excitations  out of
the Fermi sea of the leads during the resonant tunneling events. However,
we show in
the following  that such many-particle
contributions are suppressed if the resonance width
$\gamma_{l}=2\pi\nu_{l}|T_{DL}|^2$ is smaller than $\Delta\mu$
(for $\epsilon_{l}\simeq
\mu_{S}$), where $\nu_{l}$ is the density of states (DOS) per spin of
the leads at the chemical potential $\mu_{l}$.

Finally, an additional energy scale that
enters the consideration is the superconducting gap energy
$\Delta$, which is half the minimum energy it costs to break up
a Cooper pair into two quasiparticles. This gap energy also characterizes
the time
delay between the subsequent coherent Andreev tunneling events of the two
electrons of
a Cooper pair. In order to exclude single-electron tunneling where the
creation of a
quasiparticle in the superconductor is  a final excited state
we require that $\Delta >\Delta\mu, k_{B}T$.

To summarize, the regime of interest in this work is then given by
\begin{equation}
\label{regime}
\Delta, U,
\delta\epsilon>\Delta\mu>\gamma_{l},  k_{B}T,\quad
{\rm and}\quad \gamma_{l}>\gamma_{S}\, .
\end{equation}
Some inequalities will become clear when we discuss the various processes
in detail below.
As regards possible experimental implementations of the proposed setup and its
parameter regime, we would
like to mention that, typically, quantum dots are
made out of semiconducting heterostructures,
which satisfy above inequalities\cite{Kouwenhoven}.
Furthermore, in recent experiments, it has been shown
that the fabrication of hybrid structures with semiconductor and
superconductor being tunnel-coupled is   
possible\cite{{Takayanagi,Franceschi}}.
Other candidate
materials are e.g. carbon nanotubes which also show Coulomb blockade behavior
with $U$ and $\delta\epsilon$ being in the regime of interest here\cite{Dekker}.  The
present work might provide further motivation to implement the  structures 
proposed here.

Our goal in the following is to calculate the stationary charge current of pairwise
spin-entangled electrons for two competing transport channels,
first for the desired transport of two entangled electrons into
different
leads ($I_1$) and second for the unwanted transport of both electrons into the same
lead ($I_2$).
We compare then the two competing processes and show how their  ratio, $I_1/I_2$,
depends on the
various system parameters and how it can be made large. An important finding
is that
when tunneling of two electrons into different leads occurs, the current is
suppressed due to the  fact that  tunneling into the dots will typically
take place from
different points
${\bf r}_{1}$ and ${\bf r}_{2}$ on the superconductor (see Fig.\ 1) due to
the spatial
separation of
the dots $D_{1}$ and $D_{2}$. We show that the distance of separation
$\delta r=|{\bf r}_{1}-{\bf r}_{2}|$ leads to an exponential suppression of
the
current via different dots if $\delta r>\xi$ (see (\ref{I_{1}})), where
$\xi$ is the coherence length of a Cooper pair. In the relevant regime,
$\delta r<\xi$, however, the suppression is only polynomial
and $\propto 1/(k_{F}\delta r)^2$, with $k_{F}$ being the Fermi
wavevector in the superconductor. On the other hand, tunneling via
the same dot implies $\delta r=0$,
but suffers  a suppression due to $U$ and/or $\Delta$. The
suppression of this current is given by the small parameter
$(\gamma_{l}/U)^2$ in
the case $U<\Delta$, or by $ (\gamma_{l}/\Delta)^2$, if $U>\Delta$ as will
be derived in the following. Thus, to maximize the efficiency of the
entangler, we also require $k_F\delta r<\Delta, U$.

Finally, we will discuss the effect of a magnetic flux on the entangled
current in an 
Aharonov-Bohm loop, and
we will see that this current contains both, single- and two-particle Aharonov-Bohm
periods whose amplitudes have different parameter dependences.

\section{Hamiltonian of the Andreev Entangler}
\label{Hamiltonian}

We use a tunneling Hamiltonian description of the system,
$H=H_{0}+H_{T}$, where
\begin{equation}
H_{0}=H_{S}+\sum_{l}H_{Dl}+\sum_{l}H_{Ll},\qquad l=1,2.
\end{equation}
Here, the superconductor is described by the
BCS-Hamiltonian\cite{Schrieffer}
$H_{S}=\sum\limits_{{\bf k},\sigma}E_{\bf k}  \gamma_{{\bf k}\sigma}^{\dagger}
\gamma_{{\bf k}\sigma}$,
where $\sigma=\uparrow,\downarrow$, and the quasiparticle operators
$\gamma_{{\bf
k}\sigma}$ describe excitations out of the BCS-groundstate $|0\rangle_{S}$
defined by $\gamma_{{\bf k}\sigma}|0\rangle_{S}=0$. They are related to the
electron annihilation and creation operators $c_{{\bf k}\sigma}$ and
$c_{{\bf
k}\sigma}^{\dagger}$ through the Bogoliubov transformation \cite{Schrieffer}
\begin{eqnarray}
c_{{\bf k}\uparrow}&=&u_{\bf k}\gamma_{{\bf k}\uparrow}+v_{\bf
k}\gamma_{-{\bf k}\downarrow}^{\dagger}\nonumber\\
c_{-{\bf k}\downarrow}&=&u_{\bf k}\gamma_{-{\bf k}\downarrow}-
v_{\bf k}\gamma_{{\bf k}\uparrow}^{\dagger}\, ,
\label{Bogoliubov}
\end{eqnarray}
where $u_{\bf k}=(1/\sqrt{2})(1+\xi_{\bf k}/E_{\bf k})^{1/2}$ and
$ v_{\bf k}=(1/\sqrt{2})(1-\xi_{\bf k}/E_{\bf k})^{1/2}$ are the usual BCS
coherence factors \cite{Schrieffer}, and $\xi_{\bf k}=\epsilon_{\bf
k}-\mu_{S}$ is the normal state single-electron energy counted from the
Fermi level $\mu_{S}$, and $E_{\bf k}=\sqrt{\xi_{\bf k}^2+\Delta^2}$ is the
quasiparticle energy. We choose energies such that $\mu_{S}=0$. Both dots
are represented as one localized (spin-degenerate) level with energy
$\epsilon_{l}$ and is modeled by an Anderson-type Hamiltonian
$H_{Dl}=\epsilon_{l}\sum\limits_{\sigma}d_{l\sigma}^{\dagger}d_{l\sigma}+Un
_{l\uparrow}n_{l\downarrow}$, $l=1,2$. The resonant dot level $\epsilon_{l}$
can be
tuned by the gate voltage. Other levels of the dots do not participate in
transport if $\delta\epsilon>\Delta\mu>k_{B}T$,
where $\Delta\mu=-\mu_{l}$, and $\mu_{l}$ is the chemical potential of lead
$l=1,2$,  and $\delta\epsilon$ is the single particle energy level spacing
of the
dots. The leads $l=1,2$ are assumed to be non-interacting (normal) Fermi
liquids,
$H_{Ll}=\sum_{{\bf k}\sigma}\epsilon_{\bf k}a_{l{\bf
k}\sigma}^{\dagger}a_{l{\bf
k}\sigma}$. Tunneling from the dot $l$ to the lead $l$ or to the point ${\bf
r}_{l}$
in the superconductor is described by the tunnel
Hamiltonian $H_{T}=H_{SD}+H_{DL}$ with
\begin{eqnarray}
H_{SD}&=&
\sum\limits_{l\sigma} T_{SD}
d_{l\sigma}^{\dagger}\psi_{\sigma}({\bf r}_{l})+{\rm h.c.},
\label{tunnelhamilton1}\\
H_{DL}&=&
\sum\limits_{l{\bf k} \sigma} T_{DL} a_{l{\bf k}\sigma}^{\dagger}
d_{l\sigma}+{\rm h.c.}\, .
\label{tunnelhamilton2}
\end{eqnarray}
Here, $\psi_{\sigma}({\bf r}_{l})$ annihilates an electron with spin
$\sigma$ at
site ${\bf r}_{l}$,  and $d_{l\sigma}^{\dagger}$ creates it again (with the
same
spin) at dot $l$ with amplitude $T_{SD}$. $\psi_{\sigma}({\bf r}_{l})$ is
related to $c_{{\bf k}\sigma}$  by the Fourier transform $\psi_{\sigma}({\bf
r}_{l})=\sum\limits_{\bf k} e^{i{\bf kr}_{l}}c_{{\bf k}\sigma}$.
Tunneling from the
dot to the state ${\bf k}$ in the lead is described by the tunnel amplitude
$T_{DL}$.
We assume that the ${\bf k}$-dependence of  $T_{DL}$
can be safely neglected.

\section{Stationary Current and T-matrix }

The stationary current of {\it two}  electrons passing from the
superconductor
via virtual dot states to
the leads is given by
\begin{equation}
I=2e\sum\limits_{f,i}W_{fi} \rho_{i}\, ,
\label{current}
\end{equation}
where $W_{fi}$  is the transition rate from the superconductor to the
leads. We calculate this transition rate in terms of a 
T-matrix approach \cite{Merzbacher},
\begin{equation}
W_{fi}=2\pi |\langle
f|T(\varepsilon_i)|i\rangle|^2\delta(\varepsilon_f-\varepsilon_i)\,  .
\label{rate}
\end{equation}
Here,
$T(\varepsilon_i)=H_{T}\frac{1}{\varepsilon_i +i \eta-H}(\varepsilon_i-H_{0})$,
is the on-shell transmission or T-matrix,
with $\eta$ being a small positive real number which we take to zero
at the end of the calculation.
Finally,  $\rho_{i}$ is the stationary
occupation probability for the entire system to be in the state $|i\rangle$.
The T-matrix $T(\varepsilon_i)$ can be written as a power series in the
tunnel Hamiltonian $H_{T}$,
\begin{equation}
\label{series}
T(\varepsilon_i)=H_{T}+H_{T}\sum_{n=1}^{\infty}
(\frac{1}{\varepsilon_i+i\eta-H_{0}}H_{T})^n \, ,
\end{equation}
where the initial energy is $\varepsilon_i=2\mu_{S}\equiv 0$. We work in the
regime defined in Eq. (\ref{regime}), i.e. $\gamma_{l}>\gamma_{S}$, and
$\Delta,U,
\delta\epsilon>\Delta\mu>\gamma_{l}, k_{B}T $,
and around the resonance $\epsilon_{l}\simeq \mu_{S}$.
Further, $\gamma_{S}=2\pi\nu_{S}|T_{SD}|^2$ and
$\gamma_{l}=2\pi\nu_{l}|T_{DL}|^2$
define the tunneling rates between superconductor and dots, and between dots
and
leads, respectively, with $\nu_{S}$ and $\nu_{l}$ being the DOS per spin at
the chemical potentials $\mu_{S}$ and $\mu_{l}$, respectively.
We will show that the total effective tunneling
rate  from the superconductor to the leads  is given by
$ \gamma_{S}^2/\gamma_{l}$
due to the Andreev process.
In the regime
(\ref{regime}) the entire tunneling process becomes a two-particle problem
where
the many-particle effect of the reservoirs (leads) can be safely neglected
and the
coherence of an initially entangled Cooper pair (spin singlet) is maintained
during
the transport into the leads as we shall show below. Since the superconducting
gap
satisfies
$\Delta>\Delta\mu, k_{B}T$, the superconductor contains no quasiparticle
initially.
Further, in the regime (\ref{regime}), the resonant dot levels
$\epsilon_{l}$ are
mostly empty, since in the assumed asymmetric case,
$|T_{DL}|>|T_{SD}|$ (or $\gamma_{l}>\gamma_{S}$),  the electron leaves  the
dot  to
the lead much faster  than it can be replaced by another electron from the
superconductor. In addition, we can neglect any stationary  occupation of
the dots
induced by the coupling of the dots to the leads. Indeed, in the stationary
limit and
for given bias $\Delta\mu$ this occupation
probability is determined by the grand canonical distribution function
$\propto
\exp(-\Delta\mu/k_{B}T)\ll 1$,  and thus $\rho_{i}\simeq 0$ for any initial
state
where the resonant dot level is occupied. In this regime, the
initial state
$|i\rangle$ becomes
$|i\rangle=|0\rangle_{S}|0\rangle_{D}|\mu_{l}\rangle_{l}$,  where
$|0\rangle_{S}$ is the quasiparticle vacuum for the superconductor,
$|0\rangle_{D}$
means that both dot levels $\epsilon_{l}$ are unoccupied, and
$|\mu_{l}\rangle_{l}$
defines the occupation of the leads which are filled with electrons up to
the chemical potential $\mu_{l}$.
 We
remark that in our regime of interest no Kondo effects appear which could
destroy the spin entanglement, since
our dots contain each an even number of electrons in the stationary limit.

\section{Current due to tunneling into different leads}

We now calculate the current for simultaneous coherent transport of two
electrons into different leads. The final state for two electrons, one of
them being in lead 1 the other in lead 2, can be classified according to
their total spin $S$. This spin can be either a singlet (in standard
notation)
$|S\rangle=(|\!\uparrow\downarrow\rangle -
|\! \downarrow\uparrow\rangle)/\sqrt{2}$
with
$S=0$, or a triplet with $S=1$. Since the total spin is conserved, $[{\bf
S}^2,
H]=0$, the singlet state of the initial Cooper pair will be conserved in the
transport
process and the final state must also be a singlet.
That this is so can also be seen explicitly when we allow for the
possibility that
the final state could be the
$S_z=0$  triplet $|t_{0}\rangle=(|\!\uparrow\downarrow\rangle +
|\!\downarrow\uparrow\rangle)/\sqrt{2}$. [The triplets
$|t_{+}\rangle=|\!\uparrow\uparrow\rangle$ and
$|t_{-}\rangle=|\!\downarrow\downarrow\rangle$ can be excluded right away
since the
tunnel Hamiltonian $H_{T}$ conserves the spin-component
$\sigma$ and an Andreev process involves tunneling of two electrons with
different spin $\sigma$.]
Therefore we consider final two-particle states of the form
$|f\rangle=(1/\sqrt{2})[a_{1{\bf p}\uparrow}^{\dagger}a_{2
{\bf q}\downarrow}^{\dagger}\pm a_{1{\bf p}\downarrow}^{\dagger}a_{2{\bf
q}\uparrow}^{\dagger}]|i\rangle$, where the $-$ and $+$ signs belong to the
singlet $|S\rangle$ and triplet $|t_{0}\rangle$, resp. Note that this singlet/triplet
state is formed out of two electrons, one being in the  ${\bf p}$ state in lead 1
and with energy $\epsilon_{\bf p}$,
while the other one is in the ${\bf q}$ state in lead 2 with energy
$\epsilon_{\bf q}$. Thus, the two electrons are entangled in spin space
while separated in orbital space, thereby providing a non-local EPR pair.
The
tunnel process to different leads appears in the following order. A Cooper
pair breaks up, where one electron with spin $\sigma$ tunnels to one
of the
dots (with empty level $\epsilon_{l}$) from the point of the superconductor
nearest to this dot. This is a virtual
state with energy deficit $ E_{\bf k}>\Delta$. Since $\Delta>\gamma_{l}$, 
the second electron from the Cooper pair with spin $-\sigma$
tunnels to the other empty dot-level {\em before} the electron with spin $\sigma$
escapes to the
lead. Therefore, both electrons tunnel almost simultaneously to the dots
(within the uncertainty time $\hbar/\Delta$).
Since we work at the resonance $\epsilon_{l}\simeq \mu_S= 0$ the energy
denominators in
(\ref{series}) show divergences $\propto 1/\eta$ indicating that tunneling
between
the dots and the leads is resonant and we have to treat
tunneling to all orders in
$H_{DL}$ in (\ref{series}),
eventually giving a finite result in which $\eta$ will be replaced by
$\gamma_l/2$.
Tunneling back to the
superconductor is unlikely since $|T_{SD}|< |T_{DL}|$. We
can therefore write the transition amplitude between initial and final state
as
\begin{equation}
\label{amplitude1}
\langle f|T_0|i\rangle=\frac{1}{\sqrt{2}}\langle a_{2{\bf
q}\downarrow}a_{1{\bf
p}\uparrow} T^{'} d_{1\uparrow}^{\dagger}d_{2\downarrow}^{\dagger}\rangle
\langle [d_{2\downarrow}d_{1\uparrow}\pm
d_{2\uparrow}d_{1\downarrow}] T^{''}\rangle  \, ,
\end{equation}
where $T_0=T(\varepsilon_i=0)$, and 
the partial T-matrices $T^{'}$ and $T^{''}$ are given by
\begin{equation}
\label{T''}
T^{''}=\frac{1}{i\eta-H_{0}}H_{SD}\frac{1}{i\eta-H_{0}}H_{SD}\,  ,
\end{equation}
and
\begin{equation}
T^{'}=H_{DL}\sum\limits_{n=0}^{\infty} \left(\frac{1}
{i\eta-H_{0}}H_{DL}\right)^{2n+1}\, .
\end{equation}
In (\ref{amplitude1}) we used that the matrix element containing $T^{'}$ is
invariant under spin exchange $\uparrow\leftrightarrow\downarrow$, and the
abbreviation $\langle...\rangle$ stands for  $\langle i|...|i\rangle$. The part
containing $T^{''}$ describes the Andreev process, while the part containing
$T^{'}$ is the resonant dot $\leftrightarrow$  lead tunneling. 

We first
consider the Andreev process. We insert a complete set of 
single-quasiparticle (virtual) states, i.e.,
$1\!\!1=\sum_{l{\bf k}\sigma}\gamma_{{\bf
k}\sigma}^{\dagger}d_{l-\sigma}^{\dagger}|i\rangle\langle
i|d_{l-\sigma}\gamma_{{\bf k}\sigma}$, between the two $H_{SD}$ in
(\ref{T''}) and use that the resulting energy denominator $|i\eta-E_{\bf
k}-\epsilon_{l}|\approx |E_{\bf k}|$, since we work close to the resonance
$\epsilon_{l}\simeq 0$  and $E_{\bf k}>\Delta$. The triplet contribution
vanishes since $u_{\bf k}v_{\bf k}=u_{-{\bf k}}v_{-{\bf k}}$ for s-wave superconductor. 
For the
 final state being a singlet we then get
\begin{eqnarray}
&&\langle (d_{2\downarrow}d_{1\uparrow}-
d_{2\uparrow}d_{1\downarrow}) T^{''}\rangle \nonumber\\
&&\qquad\qquad =
\frac{4T_{SD}^2}{\epsilon_{1}+\epsilon_{2}-i\eta}\sum\limits_{{\bf
k}} \frac{u_{\bf k}v_{\bf k}}{E_{\bf k}}
\cos{({\bf k}\cdot\delta {\bf r})}\, ,
\label{Andreev1}
\end{eqnarray}
where $\delta {\bf r}={\bf r}_{1}-{\bf r}_{2}$ denotes the distance
vector between the points on the superconductor from which 
electron 1 and 2 tunnel into the dots.
To evaluate the sum over ${\bf k}$ we use $u_{\bf k}v_{\bf
k}=\Delta/(2E_{\bf k})$, linearize the spectrum around the Fermi
level with Fermi wavevector $k_{F}$, and obtain finally for 
the Andreev contribution
\begin{equation}
\label{Andreev2}
\langle (d_{2\downarrow}d_{1\uparrow}-
d_{2\uparrow}d_{1\downarrow}) T^{''}\rangle =
\frac{2\pi\nu_{S}T_{SD}^2}{\epsilon_{1}+\epsilon_{2}-i\eta}
\frac{\sin(k_{F}\delta r)}{k_{F}\delta r}    
e^{-\frac{\delta r}{\pi\xi}}\, .
\end{equation}

\subsection{Dominant  contribution of resonant tunneling to different leads}

Now we calculate the matrix
element in (\ref{amplitude1}) containing $T^{'}$ where tunneling has to be
treated to
all orders in $H_T$. To simplify the notation  we suppress spin indices
and
introduce a ket notation $|1 2\rangle$, where 1 stands  for quantum numbers
of the
electron on dot 1/lead 1 and similar for 2. E.g. $|p q\rangle$ stands for
$a^\dagger_{1{\bf p}\sigma} a^\dagger_{2{\bf q}-\sigma}|i\rangle$, where
${\bf p}$ is from lead 1 and ${\bf q}$ from lead 2; or $|p D\rangle$
stands for $a^\dagger_{1{\bf p}\sigma} d^\dagger_{2,-\sigma}|i\rangle$, etc.
We concentrate first on the
resummation of
the following dot $\leftrightarrow$ lead transitions $|DD\rangle\rightarrow
|LD\rangle\rightarrow|DD\rangle$ or
$|DD\rangle\rightarrow|DL\rangle\rightarrow|DD\rangle$. In this sequence,
$|DD\rangle$ is the state with electron on dot 1 and the other one on dot
2, and
$|LD\rangle$ defines a state where one electron is in lead 1 and the other
one on dot 2. We exclude processes of the kind $|DD\rangle\rightarrow
|LD\rangle\rightarrow|LL\rangle\rightarrow|LD\rangle\rightarrow|DD\rangle$
or $|DD\rangle\rightarrow
|LD\rangle\rightarrow|LL\rangle\rightarrow|DL\rangle\rightarrow|DD\rangle$,
where both electrons are {\it virtually} simultaneously in the leads as well
as the creation of electron-hole excitations out of the Fermi sea. We
show in App. A and B that such contributions are suppressed in the
regime (\ref{regime}) considered here by the small parameter $\gamma_{l}/\Delta\mu$. 
The dominant contribution is then given by 
\begin{eqnarray}
&&\langle {pq}|T^{'}|DD\rangle\hspace*{5cm} \nonumber\\
&&=\left\{
\begin{array}{r}
\langle pq|H_{D_{1}L_{1}}|Dq\rangle
\langle Dq|\sum\limits_{n=0}^{\infty}
(\frac{1}{i\eta-H_{0}}H_{D_{1}L_{1}})^{2n}| Dq\rangle\\
\times\langle Dq|\frac{1}{i\eta-H_{0}}H_{D_{2}L_{2}}|DD\rangle 
\end{array}\right.
\nonumber \\
&&\quad
\left.\begin{array}{r}
+\langle pq|H_{D_{2}L_{2}}|pD\rangle
\langle pD|\sum\limits_{n=0}^{\infty}
(\frac{1}{i\eta-H_{0}}H_{D_{2}L_{2}})^{2n}|pD\rangle\\
\times\langle pD|\frac{1}{i\eta-H_{0}}H_{D_{1}L_{1}}|DD\rangle
\end{array}
\right\}\nonumber\\
&&\times\langle DD|\sum\limits_{m=0}^{\infty}
(\frac{1}{i\eta-H_{0}}H_{DL})^{2m}|DD\rangle .
\label{resummation1}
\end{eqnarray}
Since the sums for the transition
$|DD\rangle\rightarrow|DD\rangle$  via the sequences
$|DD\rangle\rightarrow|LD\rangle\rightarrow|DD\rangle$  and
$|DD\rangle\rightarrow|DL\rangle\rightarrow|DD\rangle$
 are independent, we can write
all summations in (\ref{resummation1}) as geometric series which can be
resummed explicitly. We begin with the two-particle process for which we find
\begin{eqnarray}
&&\langle DD| \sum_{m=0}^{\infty}
(\frac{1}{i\eta-H_{0}}H_{DL})^{2m}|DD\rangle\nonumber\\
&&\qquad\qquad
=\frac{1}{1-\langle DD|
(\frac{1}{i\eta-H_{0}}H_{DL})^{2}|DD\rangle}\, ,
\label{geomseries}
\end{eqnarray}
where
\begin{equation}
\langle DD| (\frac{1}{i\eta-H_{0}}H_{DL})^{2}|DD\rangle=
\frac{\Sigma}{i\eta -\epsilon_{1}-\epsilon_{2}} \, , 
\label{selfenergy}
\end{equation}
with $\Sigma$ being the self-energy,
$\Sigma=|T_{DL}|^2\sum_{l{\bf k}}(i\eta-\epsilon_l-\epsilon_{\bf
k})^{-1}$.
In the presence of a Fermi sea in the leads, we introduce a cut-off in the sum in 
$\Sigma$
at the Fermi level $\epsilon_{\bf k}\sim -\Delta\mu$ and at the edge of the
conduction band, $\epsilon_c$. Then we obtain  
$\Sigma={\rm Re}\Sigma-i\gamma/2$,
where $\gamma=\gamma_1+\gamma_2$, and
the  logarithmic renormalization of the energy level is small, i.e. 
${\rm Re}\Sigma\sim\gamma_{l}\ln(\epsilon_{c}/\Delta\mu)\ll\Delta\mu$
and will be neglected.
Finally, we arrive at the following expression
\begin{equation}
\langle DD| \sum_{m=0}^{\infty}
(\frac{1}{i\eta-H_{0}}H_{DL})^{2m}|DD\rangle=
\frac{\epsilon_1+\epsilon_2-i\eta}{\epsilon_1+\epsilon_2-i\gamma/2}\, .
\label{resum1}
\end{equation}
Similar results hold for
the one-particle resummations in (\ref{resummation1}),  
\begin{equation}
\langle pD|\sum\limits_{n=0}^{\infty}(\frac{1}{i\eta-H_{0}}H_{D_{2}L_{2}})^{2n}|pD
\rangle= \frac{\epsilon_2+\epsilon_{\bf p}-i\eta}{\epsilon_2+\epsilon_{\bf
p}-i\gamma_{2}/2}\, , 
\label{resum2}
\end{equation}
\begin{equation}
\langle
Dq|\sum\limits_{n=0}^{\infty}(\frac{1}{i\eta-H_{0}}H_{D_{1}L_{1}})^{2n}|Dq
\rangle=\frac{\epsilon_1+\epsilon_{\bf q}-i\eta}{\epsilon_1+\epsilon_{\bf
q}-i\gamma_{1}/2}\, .
\label{resum3}
\end{equation}
Inserting the preceding results back into Eq. (\ref{resummation1}) we obtain
\begin{equation}
\label{BreitWigner1}
\langle {pq}|T^{'}|DD\rangle=\frac{-T_{DL}^2(\epsilon_{1}+\epsilon_{2}-i \eta)}
{(\epsilon_{1}+\epsilon_{\bf q}-i\gamma_{1}/2)
(\epsilon_{2}+\epsilon_{\bf p}-i\gamma_{2}/2)}\, .
\end{equation}
Thus, 
we see that the resummations cancel all divergences like the
$(\epsilon_{1}+\epsilon_{2}-i\eta)$ denominator appearing in
(\ref{Andreev1}) and (\ref{Andreev2}), and that, as expected,
the resummation of divergent terms leads effectively to the replacement
$i\eta\rightarrow i\gamma_{l}/2$ so that the limit $\epsilon_{l}\rightarrow
0$ is well-behaved. It is interesting to note that the 
 two-particle resonance
$(\epsilon_{1}+\epsilon_{2}-i\gamma/2)^{-1}$ occurring in (\ref{resum1})
has canceled out in  (\ref{BreitWigner1}), and we finally obtain  a product of two
independent single-particle  Breit-Wigner resonances. Still, we will just see
that the
two-particle correlation is reintroduced when we insert (\ref{BreitWigner1}) into
the expression for the current (\ref{current}) due to the integrations over ${\bf
p}$, ${\bf q}$,
 and the fact that the main contribution comes from the resonances.
Indeed, making use of
Eqs.\ (\ref{current},\ref{rate}), and energy conservation 
$\varepsilon_f=\varepsilon_i=0$,
i.e.
$\epsilon_{\bf p}=-\epsilon_{\bf q}$, and of Eqs. (\ref{Andreev2}) and
(\ref{BreitWigner1}),
we finally obtain for the  current (denoted by $I_{1}$)  where each of
the two entangled electrons tunnels into a {\it different} lead
\begin{equation}
\label{I_{1}}
I_{1}=\frac{e\gamma_{S}^2\gamma}{(\epsilon_1+\epsilon_2)^2+\gamma^2/4}
\left[\frac{\sin(k_{F}\delta r)}{k_{F}\delta r}\right]^2  
\exp{\{-\frac{2\delta r}{\pi\xi}\}}\, ,
\end{equation}
where, again, $\gamma=\gamma_1 +\gamma_2$. We note that Eq. (\ref{I_{1}}) also holds
for the case with $\gamma_1\neq\gamma_2$. The current becomes exponentially
suppressed with increasing distance $\delta r$ between the tunneling points on
the superconductor, the scale given by the Cooper pair coherence length $\xi$.
This does not pose severe restrictions for conventional s-wave material with
$\xi$ typically being on the order of micrometers. More severe is the restriction
that $k_{F}\delta r$ should not be too large compared to unity, especially if
$k_F^{-1}$ of the superconductor  assumes a typical value on the order of a few
Angstroms. Still, since the
suppression in $k_{F}\delta r$ is only power-law like there is  a sufficiently
large regime on the nanometer scale for
$\delta r$ where the current 
$I_{1}$ can assume a finite measurable value. The current (\ref{I_{1}}) has
again a Breit-Wigner resonance form
which assumes it maximum
value when $\epsilon_1=-\epsilon_2$,  
\begin{equation}
\label{I_{1max}}
I_{1}=\frac{4e\gamma_{S}^2}{\gamma}
\left[\frac{\sin(k_{F}\delta r)}{k_{F}\delta r}\right]^2  
\exp{\{-\frac{2\delta r}{\pi\xi}\}}\, .
\end{equation}
This resonance at $\epsilon_1=-\epsilon_2$ clearly shows that the current
is a correlated two-particle effect (even apart from any spin correlation) as we
should expect from the Andreev process involving the coherent tunneling of two
electrons.
Together with the single-particle resonances discussed above (see after Eq.\
(\ref{BreitWigner1})) we thus see that 
the current is
carried by correlated pairs of electrons whose energies satisfy 
$|\epsilon_{\bf p}| =|\epsilon_{\bf q}| \lesssim\gamma$, if $\epsilon_1 =\epsilon_2 =0$.

A particularly interesting case occurs when the energies of the dots, $\epsilon_1$
and  $\epsilon_2$, are both tuned to zero, i.e. $\epsilon_1=\epsilon_2=\mu_S=0$. We
stress that in this case the electron in lead 1 and its spin-entangled partner in
lead 2 have exactly the {\it same orbital energy}. We have shown
previously\cite{BLS} that this degeneracy  of orbital energies
is a crucial requirement
for  noise measurements in which the singlets manifest themselves in form of enhanced
noise in the current (bunching), whereas uncorrelated electrons,  or,  more generally,
electrons in a triplet state, lead to a suppression of noise (antibunching).

We remark again that the current $I_{1}$ is carried by electrons 
which are entangled in spin space and spatially separated in orbital space. In other
words, the stationary current $I_1$ is a current of non-local spin-based EPR pairs.
Finally, we note that due to the singlet character of the EPR pair we
do not know whether the electron in, say, lead 1 carries an up or a down spin,
this can be revealed only by a spin-measurement.
Of course, any measurement of the spin of one (or both) electrons 
will immediately destroy the singlet state and thus the entanglement.
Such a spin measurement (spin read-out) can  be performed e.g. by making use of the
spin filtering effect of quantum dots\cite{RSL}.
The singlet state will also be destroyed by spin-dependent scattering (but
not by Coulomb exchange interaction in the Fermi sea\cite{BLS}). However, it is 
known experimentally that
electron spins in a semiconductor environment show unusually long dephasing times
approaching microseconds and can be transported phase coherently over distances
exceeding 100$\mu m$
\cite{{Kikkawa1,Kikkawa2,Awschalom,Fiederling,Ohno}}. This distance is
sufficiently long for experiments performed typically
on the length scale of quantum confined nanostructures\cite{Kouwenhoven}.

\subsection{Negligible tunnel contributions}

We turn now to a discussion of various
 tunnel processes which we have not taken into account so far and show that
they
are negligibly small compared to the ones we have retained. As we mentioned above
we exclude virtual states where both electrons are simultaneously in the
leads. This is justified in the regime (\ref{regime}) considered here. To
show this we consider the process $|DD\rangle\rightarrow |DD\rangle$. This
transition occurs either in a transition sequence of the type
$|DD\rangle\rightarrow|LD\rangle\rightarrow|DD\rangle$, as considered above,
leading to the amplitude $A_{DL}=-i\gamma_{L}$  (see Eq. (\ref{dd}) in App. A),
or in a sequence of the type
$|DD\rangle\rightarrow|LD\rangle\rightarrow|LL\rangle\rightarrow|DL\rangle
\rightarrow|DD\rangle$, where both electrons are simultaneously in the leads
($|LL\rangle$-state), leading to the amplitude 
$A_{LL}= -\frac{\gamma_L^2}{4\pi^2\Delta\mu}\left[i\pi+\ln\left(\frac{\epsilon_c}
{\Delta\mu}\right)\right] $ (see Eq. (\ref{dlead2}) in App. A).
However,  this
amplitude $A_{LL}$ is suppressed by a factor $ \gamma_{l}/\Delta\mu<1$ compared to
$A_{DL}$. Further,
a process where we create
an electron-hole pair out of the Fermi sea of the leads could, in principle, destroy
the spin-correlation of the entangled electron pair when  an electron with the
``wrong" spin
(coming from the
Fermi sea) hops on the dot. But such contributions  cost  additional energy
of at
least
$\Delta\mu$, and again such particle-hole processes are suppressed by a factor 
$(\gamma_{l}/\Delta\mu)^2$ as we show in detail in Appendix B.

\section{Tunneling via the same dot}

The two electrons of a Cooper pair can also tunnel via the {\it same} dot into
the  same lead. In this section we calculate  the current induced by this
process. We show that we  obtain a suppression of such processes by a
factor $(\gamma_{l}/U)^2$ and/or $(\gamma_{l}/\Delta)^2$ compared to the process
discussed in the preceding section.  However, in contrast to the previous case, 
we do
not get a suppression resulting from the spatial separation of the Cooper
pair on the superconductor, since  here the two electrons tunnel from the 
same point either from ${\bf r}_{1}$ or
${\bf r}_{2}$ (see Fig.\ 2). As before, a tunnel process
starts by
breaking up a Cooper pair followed by an Andreev process with 
two possible sequences, see Fig.\ 2.
(I) In a first step, one electron tunnels from the superconductor
to, say, dot 1, and in a second step the second electron also tunnels
to dot 1. There are now 
two electrons on the {\it same} dot which  costs additional  Coulomb repulsion
energy $U$, thus this  virtual state is suppressed by $1/U$. Finally, the two
electrons leave dot 1 and tunnel into lead 1.
(II)  There is an alternative competing process (which avoids the double occupancy).
Here, one
electron tunnels to, say, dot 1, and then the same
electron tunnels  further into lead 1, leaving an excitation on the superconductor
which costs additional gap energy $\Delta$ (instead of $U$),
before finally the second electron
tunnels from the superconductor via dot 1 into lead 1.

\begin{figure}[h]
\centerline{\psfig{file=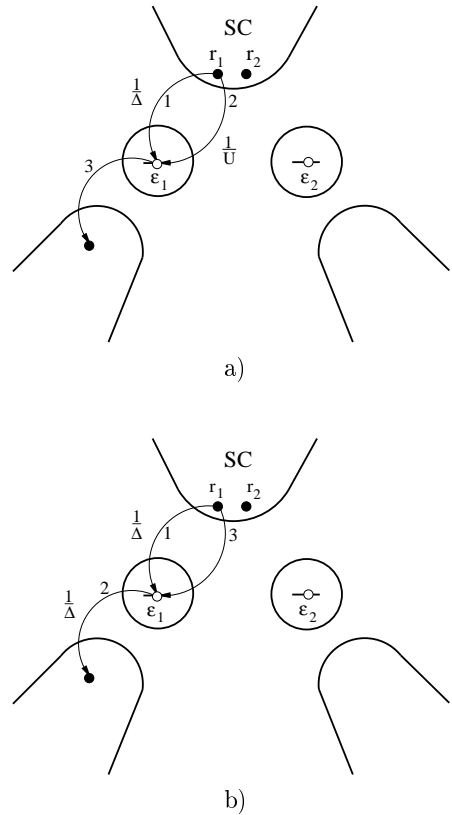,width=6cm}}
\vspace{0mm}
\caption{Two competing virtual processes are shown when the two electrons
tunnel via the same dot: 
(a)  Andreev process
leading to a double occupancy of the dot with virtual energy $ 1/U$,
and (b)
the process which differs by the sequence of
tunneling leading to an additional virtual energy $ 1/\Delta$
instead of $1/U$.}
\end{figure}

We first concentrate on the  tunneling process (II),
and note that the leading contribution comes from the processes
where both electrons have left the superconductor so that
the system has no energy deficit anymore. We still have to resum
the tunnel processes from the dot to the lead
to all orders in the tunnel Hamiltonian $H_{DL}$.
In what follows we suppress the label $l=1,2$ since the setup is assumed to be
symmetric and tunneling into either lead 1 or lead 2 gives the same result. The
transition amplitude $\langle f|T_0|i\rangle$ including only
leading terms is
\begin{eqnarray}
&&\langle f|T_0|i\rangle
=\sum\limits_{{\bf p}''\sigma}\langle
f|H_{DL}|D{\bf p}''\sigma\rangle
\nonumber\\
&&\times
\langle D{\bf p}''\sigma |
\sum\limits_{n=0}^{\infty}(\frac{1}{i\eta-H_{0}}H_{DL})^{2n}
|D{\bf p}''\sigma\rangle\nonumber\\
&&\times
\langle D{\bf p}''\sigma
|\frac{1}{i\eta-H_{0}}H_{SD}\frac{1}{i\eta-H_{0}}
H_{DL}\frac{1}{i\eta-H_{0}}H_{SD}|i\rangle\, ,
\label{series2}
\end{eqnarray}
where again $|f \rangle=(1/\sqrt{2})(a_{{\bf p}\uparrow}^{\dagger}a_{{\bf
p}'\downarrow}^{\dagger}\pm a_{{\bf p}\downarrow}^{\dagger}a_{{\bf
p}'\uparrow}^{\dagger})|i\rangle$, with $\pm$ denoting the triplet ($+$) and
singlet ($-$), resp., and the intermediate state $|D {\bf
p}''\sigma\rangle=d_{-\sigma}^{\dagger}a_{{\bf p}''\sigma}^{\dagger}|i
\rangle$. The index $\sigma$ appearing together with the tunnel Hamiltonians
in (\ref{series2}) determines the spin of the electron that tunnels. There are some
remarks in order regarding Eq. (\ref{series2}). The electron 
which tunnels to the
state $|{\bf p}''\sigma\rangle$ has not to be resummed further since this would
lead either to a double occupancy of the dot which is suppressed by $1/U$, or 
to the state with two
electrons simultaneously in the lead with a { \it virtual} summation over
the state ${\bf p''}$. But  we already showed that the latter process
is suppressed by $\gamma_{l}/\Delta\mu$.
Making then use of Eq. (\ref{resum3}),
we obtain for the first factor in (\ref{series2})
\begin{eqnarray}
&&\langle f| H_{DL}\sum\limits_{n=0}^{\infty}
(\frac{1}{i\eta-H_{0}}H_{DL})^{2n}|D{\bf p}''\uparrow\rangle
\nonumber\\
&&\qquad\qquad
= -\frac{T_{DL}}{\sqrt{2}}
\frac{\epsilon_{l}+\epsilon_{{\bf
p}''}-i\eta}{\epsilon_{l}+\epsilon_{{\bf
p}''}-i\gamma_l/2}
\left(\delta_{{\bf p}''{\bf p}}\mp\delta_{{\bf p}''{\bf p}'}\right)\, ,
\label{uparrow}
\end{eqnarray}
\begin{eqnarray}
&&\langle f| H_{DL}\sum\limits_{n=0}^{\infty}
(\frac{1}{i\eta-H_{0}}H_{DL})^{2n}|D{\bf p}''\downarrow\rangle
\nonumber\\
&&\qquad\qquad
=\frac{T_{DL}}{\sqrt{2}}
\frac{\epsilon_{l}+\epsilon_{{\bf
p}''}-i\eta}{\epsilon_{l}+\epsilon_{{\bf
p}''}-i\gamma_l/2}
\left(\delta_{{\bf p}''{\bf p}'}\mp\delta_{{\bf p}''{\bf p}}\right)\, ,
\label{downarrow}
\end{eqnarray}
where again in (\ref{uparrow}) and (\ref{downarrow}) the upper sign belongs
to the triplet and the lower sign to the singlet. For the third line of
(\ref{series2}) containing the superconductor-dot transitions we
obtain
\begin{eqnarray}
&& \langle D{\bf
p}''\uparrow|\frac{1}{i\eta-H_{0}}H_{SD}\frac{1}{i\eta-H_{0}}H_
{DL}\frac{1}{i\eta-H_{0}}H_{SD}|i\rangle 
\nonumber\\
&&=-
\langle D{\bf
p}''\downarrow|\frac{1}{i\eta-H_{0}}H_{SD}\frac{1}{i\eta-H_{0}}H_
{DL}\frac{1}{i\eta-H_{0}}H_{SD}|i\rangle
\nonumber\\
&&
 =\frac{T_{DL}T_{SD}^2\nu_{S}}{\Delta(\epsilon_{l}+\epsilon_{{\bf
p}''}-i\eta)}.
\label{downarrow2}
\end{eqnarray}
Combining the results  (\ref{uparrow})-(\ref{downarrow2}) we obtain for
the amplitude (\ref{series2})
\begin{equation}
\langle
f|T_0|i\rangle=
-\frac{2^{3/2}\nu_{S}(T_{SD}T_{DL})^2
(\epsilon_{l}-i\gamma_{l}/2)}
{\Delta(\epsilon_{l}+\epsilon_{\bf p}-i\gamma_{l}/2)
(\epsilon_{l}+\epsilon_{{\bf p}'}-i\gamma_{l}/2)}
\label{delta}
\end{equation}
for the final state $|f\rangle$ being a singlet, whereas we get again zero for the
triplet. 

Next we consider the process (I) where the tunneling involves a double occupancy
of the dot (see left panel in Fig.\ 2). In this case the transition
amplitude can be written as
\begin{eqnarray}
&&\langle f|T_0|i\rangle
=\sum\limits_{{\bf p}''\sigma}\langle f|H_{DL}|D{\bf p}''\sigma\rangle
\nonumber\\
&&\times
\langle D{\bf p}''\sigma|\sum\limits_{n=0}^{\infty}
(\frac{1}{i\eta-H_{0}}H_{DL})^{2n}|D{\bf
p}''\sigma\rangle\nonumber\\
&&\times\langle D{\bf p}''\sigma|\frac{1}{i\eta-H_{0}}H_{DL}
\frac{1}{i\eta-H_{0}}H_{SD}\frac{1}{i\eta-H_{0}}H_{SD}|i\rangle\,.
\label{series3}
\end{eqnarray}
As before, the transition amplitude $\langle
f|T_0|i\rangle$ is only nonzero for the final lead state $|f\rangle$ being
a singlet
state. Repeating a similar calculation as before we find that the amplitude
is given by (\ref{delta}) but with $\Delta$ being replaced by $U/\pi$.
We note that the two amplitudes (\ref{delta}) and (\ref{series3})
have the same initial and same final states. Thus, to obtain the total current
due to processes (I) and (II) we need to add these two amplitudes.
Then, using Eq.\ (\ref{current}) we find for the total current $I_2$
in case of tunneling of two electrons into the same lead,
\begin{equation}
I_{2}=\frac{2e\gamma_S^2\gamma}{{\cal E}^2},\qquad
\frac{1}{\cal E}=\frac{1}{\pi\Delta}+\frac{1}{U}\, .
\label{I2}
\end{equation}
We see that the effect of the quantum dots consists in the suppression
factor $(\gamma/{\cal E})^2$ for tunneling into the {\it same} lead.
We remark that in contrast to the previous case (tunneling into different
leads) the current
does not have a resonant behavior since the virtual dot states are no longer at
resonance due the energy costs $U$ or $\Delta$ in the tunneling process.
Our final
goal is to compare $I_{1}$ given in (\ref{I_{1max}}) with $I_{2}$. Thus, forming the
ratio  of the currents of the 
two competing processes, we obtain
\begin{equation}
\label{final}
\frac{I_{1}}{I_{2}}=
\frac{2{\cal E}^2}{\gamma^2}
\left[\frac{\sin(k_{F}\delta r)}{k_{F}\delta r}\right]^2   \exp\{-\frac{2\delta
r}{\pi\xi}\}\, .
\end{equation}
From this ratio we see that the desired regime  with $I_1$ dominating
$I_2$ is obtained when ${\cal E}/\gamma > k_F\delta r$, and $\delta r<\xi$.
We would like to emphasize that the relative
suppression of $I_2$ (as well as the absolute value of the current $I_1$) is
maximized by working around the resonances $\epsilon_{l}\simeq \mu_S= 0$\cite{incoherent}.

\section{Discussion and Aharonov-Bohm oscillations }

We have seen that there are two competing processes of currents, 
one where the two
electrons proceed via different dots into different leads, and one where the two
electrons proceed via the same dot into the same lead. We will show now  that these
two processes also lead to different  current oscillations in an Aharonov-Bohm loop
which is threaded by an external
magnetic flux $\phi$. For this let us
consider now a setup where the two leads 1 and 2
are connected  such that they form an Aharonov-Bohm loop,
where the electrons are injected from the left via the superconductor, traversing
the upper (lead 1) and lower (lead 2) arm of the loop before they rejoin
to interfere and then exit into the same lead, where the current is then
measured as a function of varying flux $\phi$. 
It is straightforward to analyze this setup
with our results obtained so far. In particular, each tunneling amplitude
obtains a phase factor, $T_{D_1L_1}\rightarrow  T_{D_1L_1} e^{ i\phi/2\phi_0}$,
and $T_{D_2L_2}\rightarrow  T_{D_2L_2} e^{ -i\phi/2\phi_0}$,
where $\phi_0=h/e$ is the single-electron flux quantum. 
For simplicity we also assume that the entire 
phase is acquired  when the electron hops from the dot into the leads, so that
the process dot-lead-dot gives basically the full Aharonov-Bohm
phase factor $e^{\pm i\phi/\phi_0}$ of the loop (and only a negligible  amount of
phase is picked up
along the path from the superconductor to the dots). 
Now, we repeat
the calculations of the transition amplitude   
and find it to be of the following structure
$\langle f|T_0|i\rangle \sim 
T_{D_1L_1} T_{D_2L_2}+T_{D_1L_1}^2e^{ i\phi/\phi_0} +  
T_{D_2L_2}^2e^{ -i\phi/\phi_0}$.
Here, the first term comes from the process via different leads (see
(\ref{BreitWigner1})), where no Aharonov-Bohm phase is picked up. 
The Aharonov-Bohm phase appears in the 
remaining two terms, which come from processes via the same leads, either via lead
1 or lead 2 (see (\ref{delta}) and (\ref{series3})). 
The total current $I$ is now obtained from $|\langle f|T_0|i\rangle|^2$, giving
$I=I_1 +I_2 +I_{AB}$, and the
flux-dependent Aharonov-Bohm current $I_{AB}$
is given by
\begin{eqnarray}
I_{AB}=   \sqrt {8I_1 I_2}F(\epsilon_l) 
\cos {(\phi/\phi_0)} +  I_2 \cos {(2\phi/\phi_0)}, 
&&
\label{ABcurrent}\\
F(\epsilon_l)=\frac{\epsilon_l}{\sqrt{\epsilon_l^2+(\gamma_L/2)^2}},
&&
\label{F-factor}
\end{eqnarray}
where, for simplicity, we have assumed that $\epsilon_1=\epsilon_2=\epsilon_l$,
and $\gamma_1=\gamma_2=\gamma_L$.
Here, the first term (different leads) is periodic in $\phi_0$ like
for  single-electron Aharonov-Bohm interference effects, while the second one
(same leads) is periodic in {\it half} the flux quantum $\phi_0/2$, describing
thus the interference of two coherent electrons (similar single- and
two-particle Aharonov-Bohm effects occur in the
Josephson current through an Aharonov-Bohm loop\cite{CBL}). It is clear from
(\ref{ABcurrent}) that the $h/e$ oscillation
comes from the interference between a contribution where the two electrons travel
through different arms with  contributions where the two electrons travel through the
same arm. 
Both Aharonov-Bohm oscillations with period
$h/e$, and
$h/2e$, vanish with decreasing $I_2$, i.e. with
increasing on-site repulsion $U$ and/or gap $\Delta$.
However, their relative weight is given by $\sqrt{I_1/I_2}$, implying
that the $h/2e$ oscillations vanish faster than the $h/e$ ones. This
behavior is quite remarkable since it opens up the possibility to 
tune down the unwanted
leakage process $\sim I_2 \cos {(2\phi/\phi_0)}$ where two electrons proceed via the
same dot/lead by  increasing $U$ with a gate voltage applied to the dots.
The dominant current contribution with period $h/e$ comes then from the desired
entangled electrons proceeding via different leads.
On the other hand, if $\sqrt{I_1/I_2}<1$, 
which could become the case e.g. for $k_F\delta r>{\cal E}/\gamma$, we are left with  $h/2e$
oscillations only.
Note that dephasing processes which affect the orbital part suppress $I_{AB}$.
Still, the flux-independent current $I_1 + I_2$ can remain finite and  contain
electrons which are entangled in spin-space, provided that there is
only negligible spin-orbit coupling so that the spin is still a good
quantum number.

We would like to mention another important feature of the Aharonov-Bohm
effect under discussion, namely the relative phase shift between the
amplitudes of tunneling to the same lead and to different leads, resulting in the
additional prefactor $F(\epsilon_l)$ in the first term of the rhs.\ of the Eq.\
(\ref{ABcurrent}). This phase shift is due to the fact that there is a two-particle
resonance in the amplitude (\ref{BreitWigner1}) while there is only a single-particle
resonance in the amplitudes (\ref{delta}) and (\ref{series3}) (we recall that the second
resonance is suppressed by the Coulomb blockade effect). Thus, when the chemical potential
$\mu_{S}$ of the superconductor crosses the resonance, 
$|\epsilon_l|\lesssim\gamma_L$, the amplitude
(\ref{BreitWigner1}) acquires an extra phase factor $e^{i\phi_r}$,
where $\phi_r=\arg[1/(\epsilon_l-i\gamma_L/2)]$. Then the interference
of the two amplitudes leads to the prefactor $F(\epsilon_l)=\cos\phi_r$
in the first term in the rhs of (\ref{ABcurrent}). In particular,
exactly at the middle of the resonance, $\epsilon_l=0$, the phase shift
is $\phi_r=\pi/2$, and thus the $h/e$ oscillations vanish, since $F(0)=\cos(\pi/2)=0$.
Note however, that although $F=\pm 1$ away from the resonance 
($|\epsilon_l|\gg\gamma_L$) the $h/e$ oscillations vanish again,
now because the current $I_1\sim e\gamma_S^2\gamma_L/\epsilon_l^2$ vanishes.
Thus the optimal regime for the observation of the Aharonov-Bohm effect 
is $|\epsilon_l|\sim\gamma_L$.

Finally, the preceding discussion shows that  even if the spins of two
electrons are entangled
their associated charge current does not reveal this spin-correlation
in a simple Aharonov-Bohm interference experiment\cite{ABdoubledot}. 
Only if we consider the current-current correlations (noise) in a beam splitter setup,
can we detect also this spin-correlation in the transport current via its charge
properties\cite{BLS}.

\section{Conclusion}

We have proposed  an entangler device that can create 
pairwise spin-entangled electrons and provide  coherent injection
by an Andreev process
into different dots which are tunnel-coupled to leads. 
The unwanted process of both electrons
tunneling into the same leads can be suppressed by increasing 
the Coulomb repulsion on the quantum dot. We have calculated the ratio
of currents of these two competing processes and shown that
there exists a regime of experimental interest where the entangled
current shows a resonance and assumes
a finite value with both partners of the singlet being in different
leads but having the same orbital energy. This entangler then satisfies
the necessary requirements needed to detect the spin entanglement 
via transport and noise measurements. We also discussed the 
flux-dependent 
oscillations  of the current in an Aharonov-Bohm loop.

After finishing this project we learned of a
recent preprint by Lesovik et al.\cite{Lesovik} which also makes use of the
Andreev process to generate spin-entangled electrons in the normal regime.
There, the electrons are assumed to tunnel into   fork-shaped normal leads
with no Coulomb blockade behavior.
The  separation of the entangled electrons is achieved via
energy filters so that
the two electrons enter their corresponding
lead with different orbital energies (while in the setup proposed
here the energies can be made equal).

\acknowledgments
We would like to thank Guido Burkard and Gianni Blatter for useful
discussions.
This work has been supported by the Swiss National Science Foundation.

\appendix
\section{suppression of virtual states with both electrons in the
leads}
We have stated in the main text that the contributions of virtual states
where
two electrons are
simultaneously in the leads  are negligible.
Here we estimate this contribution and show that indeed it is
suppressed by $\gamma_{L}/\Delta\mu\ll 1$
(here the spin of the electrons is not important,
and we set $\gamma_1=\gamma_2=\gamma_L$ for simplicity).
First we consider the dominant transition  from $|DD\rangle$ back to
$|DD\rangle$ with the tunneling of only one electron to the lead,
i.e.\ a sequence of the type $|DD\rangle\rightarrow|LD\rangle\rightarrow|DD\rangle$,
and we find for the amplitude (cf. Eqs. (\ref{geomseries},\ref{selfenergy}))
\begin{equation}
A_{DL}=\langle DD|H_{DL}\frac{1}{i\eta-H_{0}}H_{DL}|DD\rangle
=-i\gamma_{L}\, .
\label{dd}
\end{equation}
We compare this amplitude $A_{DL}$ with the amplitude $A_{LL}$ of the lowest-order
process
of tunneling of two electrons via the virtual
state  $|LL\rangle$, where both electrons are simultaneously in the 
leads,
i.e. the sequence
$|DD\rangle\rightarrow|LD\rangle\rightarrow|LL\rangle\rightarrow|DL\rangle
\rightarrow|DD\rangle$. From now on  we impose the resonance condition
$\epsilon_{l}=0$, and 
we find 
\begin{eqnarray}
&&A_{LL}=
\langle DD|H_{DL}\left(\frac{1}{i\eta-H_{0}}H_{DL}\right)^3
|DD\rangle 
\nonumber\\
&&=\sum\limits_{{\bf k}{\bf k}'}\frac{|T_{DL}|^4}{(i\eta-\epsilon_{\bf k}-
\epsilon_{{\bf k}'})(i\eta-\epsilon_{\bf k})}
\left[\frac{1}{i\eta-\epsilon_{\bf k}}+\frac{1}{i\eta-\epsilon_{{\bf
k}'}}\right],
\label{dlead1}
\end{eqnarray}
where the first term in the bracket results from the sequence of, say,
electron 1 tunneling into lead 1, then electron 2 tunneling into
lead 2, then electron 2 tunneling back into dot 2, and finally
electron 1 tunneling back into dot 1. While the second term in the bracket
results from the sequence where the order of tunneling back to the dots
is reversed, i.e. electron 1 tunnels  back to its dot before electron 2 does.
Note that due to this two terms in the bracket the two-particle pole in (\ref{dlead1})
cancels.
 
Replacing $\sum_{{\bf k}}(\ldots)$
with $\nu_L\int_{-\Delta\mu}^{\epsilon_{c}}d\epsilon(\ldots)$,
we can write
\begin{eqnarray}
A_{LL}
&&=\frac{\gamma_{L}^2}{(2\pi)^2}
\int\limits_{-\Delta\mu}^{\epsilon_{c}}\frac{d\epsilon'}{i\eta-\epsilon'}
\int\limits_{-\Delta\mu}^{\epsilon_{c}}\frac{d\epsilon}{(i\eta-\epsilon)^2}
\nonumber\\
&&=-\frac{\gamma_L^2}{4\pi^2\Delta\mu}\left[i\pi+\ln\left(\frac{\epsilon_c}
{\Delta\mu}\right)\right].
\label{dlead2}
\end{eqnarray}
Thus, comparing $A_{DL}$  with $A_{LL}$,  
we see that indeed a virtual
state involving two electrons simultaneously in the leads is suppressed by a
factor of $\gamma_{L}/\Delta\mu$ compared to the one with only one
electron in the leads.

\section{electron hole pair excitation}

In this Appendix we consider a tunnel process where the two electrons
starting from the superconductor tunnel over different dots but during the
process of repeated tunneling  from the dots to the leads and back to the dots
an electron from
the Fermi sea hops on one of the dots (say dot 1) when this dot is empty.
In principle,
such contributions could destroy the desired entanglement since then a
``wrong" spin
can hop on the dot and the electron on the other dot (dot 2) would no longer
be entangled
with this electron (while the original partner electron disappears in the
reservoir
provided by the Fermi sea).
We show now that in the regime
$\Delta\mu>\gamma_{l}$ such electron-hole pair processes due to the Fermi
sea are
suppressed. We start with our consideration when the two electrons,
after the Andreev process, are each on a different dot forming the
$|DD\rangle$-state (we neglect spin in this consideration for simplicity).
Instead of
the amplitude
$\langle pq|T'|DD\rangle$ calculated in (\ref{resummation1}) we consider now
the
following process
\begin{eqnarray}
A_{eh}
&&=\langle\overline{pq}|T'|\overline{DD}\rangle
\nonumber\\
&&\times
\left\{
\begin{array}{r}
\langle\overline{DD}|\frac{1}{i\eta-H_{0}}H_{D_{1}L_{1}}
\sum\limits_{n=0}^{\infty}
(\frac{1}{i\eta-H_{0}}H_{D_{2}L_{2}})^{2n}\\
\times\frac{1}{i\eta-H_{0}}H_{D_{1}L_{1}}|DD\rangle 
\end{array}
\right\}
\nonumber\\
&&\times
\langle
DD|\sum\limits_{m=0}^{\infty} 
(\frac{1}{i\eta-H_{0}}H_{DL})^{2m}|DD\rangle\, .
\label{resummation3}
\end{eqnarray}
The  new sequence of interest in (\ref{resummation3}) is the amplitude
containing the sum over $n$. For instance, let us consider the $n=0$ term,
$\langle\overline{DD}|(\frac{1}{i\eta-H_{0}}H_{D_{1}L_{1}})^2
|DD\rangle$, where we assume that the
electron-hole
excitation occurs in, say, lead 1. From $|DD\rangle$, the tunnel Hamiltonian
$H_{D_{1}L_{1}}$ takes the electron from dot 1 to the state ${\bf k}$ in lead 1.
Instead of tunneling back of this electron to  dot 1, an electron from
the state  ${\bf k}'$  with energy $\epsilon_{{\bf k}'}<-\Delta\mu$ from the
Fermi sea of lead 1 hops on dot 1. Now the dot-lead system is in the state
$|\overline{DD}\rangle=d_{1}^{\dagger}d_{2}^{\dagger}a_{1{\bf k}'}
a_{1\bf k}^{\dagger}|i\rangle$. The sum over $n$ resums the hoping back and
forth of electron 2 from $D_2$ to $D_2$, resulting in the replacement
of  $\eta$ in $H_{D_{1}L_{1}}(i\eta-H_{0})^{-1}H_{D_{1}L_{1}}$ by $\gamma_L/2$. 
We perform the further resummation in (\ref{resummation3})  with
this
Fermi sea electron on dot 1 and the other electron still on dot 2, assuming that 
electron 1 in the  state ${\bf k}$ in lead 1 is in its final state (and not a virtual
state). 
All the resummation processes in (\ref{resummation3}) are similar to the ones
already explained in the main text, except for having now an excitation with energy
$\epsilon_{\bf k}-\epsilon_{{\bf k}'}>0$. 
The final state
$|\overline{pq}\rangle$ consists of two electrons in the lead states
${\bf p}$ and ${\bf q}$ (their multiple tunneling is resummed in $T'$) and of the
excitation with energy
$\epsilon_{\bf k}-\epsilon_{{\bf k}'}$, 
so $|\overline{pq}\rangle=a_{1{\bf p}}^{\dagger}a_{2{\bf
q}}^{\dagger}a_{1{\bf k}'}a_{1{\bf k}}^{\dagger}|i\rangle$.
The normalized correction to the current, $I_{eh}/I_1$, can be obtained
by summing $|A_{eh}|^2/I_1$ over the 
 final states $|\overline{pq}\rangle$, 
and thus we arrive at the following integral
for $\epsilon_l=0$, retaining only leading terms in $\gamma_L/\Delta\mu$,
and using energy conservation,
$\epsilon_{\bf k}-\epsilon_{{\bf k}'}+\epsilon_{\bf p}+\epsilon_{\bf q}=0$,
\begin{eqnarray}
\frac{I_{eh}}{I_1}
&&=\left(\frac{\gamma_{L}}{2\pi}\right)^3
\int\!\!\!\!\int\limits_{-\Delta\mu}^{+\infty}\!\!\!\!\int 
d\epsilon_{\bf k}d\epsilon_{\bf p}d\epsilon_{\bf q}
\nonumber\\
&&\times
\frac{1-\theta(\epsilon_{\bf k}+\epsilon_{\bf p}+\epsilon_{\bf q}+\Delta\mu)}
{[\epsilon_{\bf k}^2+(\gamma_{L}/2)^2]
[\epsilon_{\bf p}^2+(\gamma_{L}/2)^2]
[\epsilon_{\bf q}^2+(\gamma_{L}/2)^2]}\, .
\label{A2}
\end{eqnarray}
We evaluate the integral in leading order and find
\begin{equation}
\frac{I_{eh}}{I_1}=\frac{3}{2\pi^2}
\left(\frac{\gamma_L}{\Delta\mu}\right)^2
\log\left(\frac{\Delta\mu}{\gamma_{L}}\right).
\end{equation}
We see now that the current involving an electron-hole pair, $I_{eh}$, is
suppressed
compared to the main contribution $I_{1}$ (see (\ref{I_{1}})) by a factor of
$(\gamma_{L}/\Delta\mu)^2$.


\begin{references}
\bibitem{ABdoubledot}
We note, however, that the Aharonov-Bohm current can be used as a probe to detect
localized spin-singlets in coupled double-dots\cite{LS}.
\bibitem{incoherent}
We remark that incoherent
transport (sequential tunneling) is negligible as long as the scattering
rate inside the dots, $\Gamma_{\varphi}$, is much smaller than $\gamma_{l}$, since
$I_{{\rm seq}}/I_{{\rm coh}}\simeq \Gamma_{\varphi}/\gamma_{l}$
\cite{Datta}.
\end{references}
\end{document}